# A MULTI-RELATIONAL NETWORK TO SUPPORT THE SCHOLARLY COMMUNICATION PROCESS[1]

MARKO A. RODRIGUEZ
Digital Library Research & Prototyping Team
Los Alamos National Laboratory
Los Alamos, New Mexico 87545
marko@lanl.gov

**Abstract**
The general purpose of the scholarly communication process is to support the creation and dissemination of ideas within the scientific community. At a finer granularity, there exists multiple stages which, when confronted by a member of the community, have different requirements and therefore different solutions. In order to take a researcher's idea from an initial inspiration to a community resource, the scholarly communication infrastructure may be required to 1) provide a scientist initial seed ideas; 2) form a team of well suited collaborators; 3) locate the best venue to publish the formalized idea; 4) determine the most appropriate peers to review the manuscript; and 5) disseminate the end product to the most interested members of the community. Through the various delineations of this process, the requirements of each stage are tied solely to the multi-functional resources of the community: its researchers, its journals, and its manuscripts. It is within the collection of these resources and their inherent relationships that the solutions to the stages of scholarly communication are to be found. This paper describes a multi-relational network composed of multiple scholarly artifacts that can be used as a medium for supporting the scholarly communication process.

## 1. Introduction

The scholarly communication process transforms the vague ideas of a researcher into a formalized paper ready for use by the scientific community. This transformation process is composed of multiple individual stages that each has their unique requirements and therefore their unique solutions. The particular stages of this process may include, but are definitely not limited to, 1) finding papers related to one's idea, 2) finding collaborators to work on the idea with, 3) finding a public venue to publish the formalized idea, 4) finding referees to review the submitted paper, and 5) finally finding the most interested members of the community to read the published paper. It is obvious that the solutions to these problems remain tied solely to scholars, papers, and journals. Therefore, for each of these stages, problem solving is a matter of searching the pool of scholarly resources for the most appropriate solution set given the particular problem context.

---

[1] Rodriguez, M.A., "A Multi-Relational Network to Support the Scholarly Communication Process", International Journal of Public Information Systems, volume 2007, issue 1, ISSN: 1653-4360, 2007.

No scholarly resource is an island unto itself and the scholarly record makes the relationship between resources explicit. Ontologically, an author writes a paper, a paper cites another paper, a paper is published in a journal, and a journal is associated with a particular scholarly domain. Given any of the stages of the scholarly communication process, individuals pursue the chain of associations that bring them from one scholarly resource to another until they arrive at a resource, or set of resources, which solves their particular problem at hand. For instance, when a researcher wants to find papers related to an inspiring paper they have read, they may spend their time looking at the references of the paper, perusing the journal associated with the paper, and scanning the paper's author's websites for other publications by the author. Likewise, journal editors may utilize a paper's references and past publishing authors of the journal as potential referees for peer-review.

Currently, many computational systems exist that support the locating of scholarly papers. Google Scholar[2] is a relatively recent development for locating scholarly papers that works in a manner similar to the Google web search engine. Google Scholar allows users to search by keyword matching and/or moving through the paper citation network. The keyword and citation network mechanism is a popular interface and is the primary search mechanism for many other repositories such as Citebase[3] (an extension to ArXiv[4]), Citeseer[5], and the many publisher web interfaces. Another mechanism that has become popular in recent years is the collaborative tagging method [Golder & Huberman, 2006]. CiteULike[6] allows registered users to add 'tags' (keyword metadata) to papers in order to provide a keyword/user search environment. People can search for papers by tags or by looking at the papers their favorite 'taggers' have tagged. ArXiv also supports push-based recommendation of scholarly papers. ArXiv users can register their email and their favorite scholarly domains and every U.S. east coast morning, an email is sent to them containing the recently submitted and update papers associated with their favorite domains. Finally, usage-networks have been employed as a means of finding related resources in a scholarly repository [Bollen & Van de Sompel, 2006] in a manner similar to the collaborative filtering algorithms of recommender systems [Herlocker & et.al., 2004].

Scholars use these systems to locate resources. However, the scholarly communication process is not simply about finding papers, but also about finding collaborators, publication venues, peer-reviewers, and an audience for one's ideas. As will be presented in this paper, both the problems and the solutions of scholarly communication can be represented as a collection of resources within a digital-library repository, where problem solving is the act of mapping one collection of resources to another. Figure 1 depicts the stages of the scholarly communication process and the mapping of problems to solutions in this process.

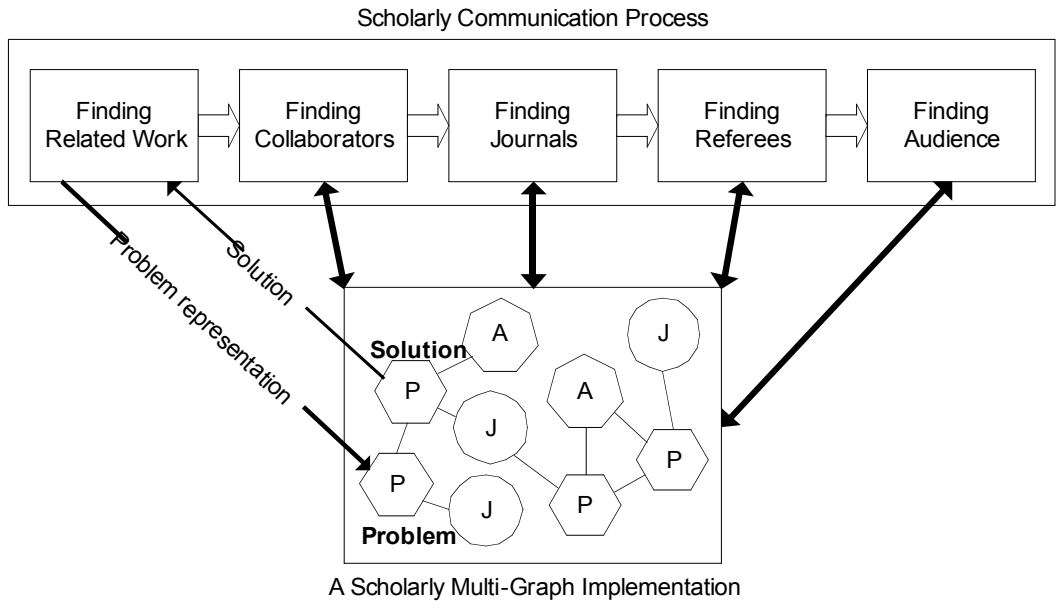

Figure 1. Problems in the scholarly communication process and how they are represented and solved by the same network substrate (A: author, P: paper, J: journal)

The five problem solving processes solved by the proposed system are:

1. *locating references for an initial idea*: the system determines a collection of manuscripts related to an individual's rough and unformulated idea.

2. *finding potential collaborators for an idea*: given a model of the specific domain of the individual's initial idea, system will suggest fit collaborators.

3. *determining a journal for paper submission*: once the collaborating team has formalized the idea into a written manuscript, a journal suitable for the paper will be recommended by the system.

4. *locating peer-reviewers to review a paper*: journal editors can use the system to find qualified referees to review the submitted manuscript.

5. *distribute the accepted paper to appropriate members in the community*: the system creates a plausible mapping between the newly published paper and potentially interested scientists in the field.

Unlike modern scholarly paper search engines, the proposed system does not simply provide a substrate for perusing the citation network of papers or matching keywords. Instead, the proposed system simulates the process of scholars searching the scholarly network in the more 'manual' association-based search. This simulated search identifies resources related to the scholar's problem. The behavior of a scholar utilizing the web of associations between various scholarly artifacts can be simulated in a manner similar to the random walk algorithm of Markov chain analysis [Häggström, 2002], the continuous-form spreading activation algorithms of information retrieval [Crestani, 1997b], and by a non-converging process similar to the random surfer model of the PageRank algorithm developed by Google [Brin & Lawrence, 1998]. In short, a random walker moves from node to node in the scholarly network identifying the most related artifacts to one's problem as being those that are most visited by the random walker. While not discussed in this article, a more advanced constrained random walker model can be used called the grammar-based random walker model of network analysis [Rodriguez, 2007]. In grammar-based simulated search model, the random walker can only traverse particular edges such that the random walker's path is limited to a subset of the full scholarly network.

The primary purpose of this paper is to describe a single unified system that simulates the search processes of scholars during the five stages of the scholarly communication process—a multi-artifact recommender system. Many scholarly repositories contain well-maintained metadata for their resources. Unlike the World Wide Web, where data is poorly structured and metadata is difficult to obtain, the metadata in a scholarly repository is much richer. This metadata can be used in novel ways to facilitate searching for resources whether those resources are authors, papers, or journals and conference proceedings and can be used beyond keyword searches. While keywords searches have shown their usefulness, other technologies such as the technology proposed by this paper can be used to augment the search capabilities of modern scholarly repositories. The proposed system is not intended to replace existing search technologies, but instead to provide a complementary data structure and algorithm that takes novel advantage of the data stored in scholarly repositories.

## 2. A Multi-Relational Scholarly Network

A multi-relational network is a network composed of a set of heterogeneous artifacts (nodes) connected to one another by a set of heterogeneous relationships (edges). Two nodes can be connected to one another by more than one edge. Thus, in order to make a distinction between two edges between the same two nodes, a label is used. A label qualifies the meaning of an edge and denotes the type of relationship between the two nodes. For instance, an author can be related to another because they are both part of the same institution and/or because they have co-authored a paper together. Furthermore, a multi-relational network can be weighted such that the association between two nodes is 'stronger' if their weight is higher. For the remainder of this paper, a multi-relational is defined by the tuple $G = (N, E, W)$, where $N$ is the set of nodes in the network, $E$ is a set of directed edges, $W$ is the set of weights associated with each edge of the network, $E \subseteq N \times N$, and $|W| = |E|$. Any edge, $e_{i,j}^\mu$, and associated weight, $w_{i,j}^\mu$, represent a directed relationship from node $n_i$ to node $n_j$ created by means of metadata of type $\mu$.

Therefore, unlike the World Wide Web, where an edge is simply a directed weightless relationship that states that node $A$ references node $B$, a multi-relational network can represent a heterogeneous population of nodes and a semantically rich set of edges.

## 2.1. Repository Metadata Records Stored at LANL

Many institutions utilize digital-libraries to electronically publish various digital-objects associated with scholarly communication. Digital-library technology not only stores digital-objects, but also maintains a record of each resource's associated metadata. Of particular interest to this paper's presented system are the digital-objects representing electronic manuscripts—pre-prints and published papers. Manuscript metadata such as authoring scientists, citation references, and publishing journals make it possible to algorithmically generate a portion of the scientific community's multi-relational network model. This section will describe how LANL's digital-library metadata is used to construct a multi-relational network of authors, papers, and journals/conferences. Given the current standards in metadata representation and harvesting, a similar paradigm can be used to harvest data from public repositories such as ArXiv and CiteSeer.

The most widely accepted digital-library metadata harvesting protocol is the OAI-PMH[7]. OAI-PMH has been used by institutional repositories worldwide as a gateway for accessing the metadata associated with repository resources. When accessing a particular resource's metadata record by way of OAI-PMH, an XML representation of the metadata is returned within the high-level `<metadata>` tag. The `<metadata>` tag contains all the metadata associated with a particular repository resource. A repository record can encode its metadata in one or more metadata formats such as Dublin Core[8] or MARC[9]. Example 1 is a particular metadata record for a manuscript stored at the LANL Research Library. The record's metadata is displayed according to the DIDL (`<didl:>`) namespace [Bekaert et al., 2003]. For the sake of brevity, multiple tags that do not pertain to the goals of this paper have been left out. Also, the references section has been shortened. Important tags that will be noted in the following subsections are highlighted with bold font and have the **Δ** notation near the right hand margin. The **Δ#** notation refers to metadata necessary to generate intra-layer connections and **Δ#-#** refers to metadata necessary to create inter-layer connections of the multi-relational network.

```
<record>
  <header>
    <identifier>info:lanl-repo/i/0e08eefc-d053-11d8-85e1-
d1cbfd475562</identifier>
    <datestamp>2004-07-07T20:20:14Z</datestamp>
    <setSpec>format:info*3Alanl-repo*2Fpro*2Fisi</setSpec>
  </header>
  <metadata>
    <didl:Resource mimeType="application/isi+xml">
```

```
<citation>
  <title>The meaning of self-organization in computing</title>
  <author>
    <au>Heylighen, F</au>                              Δ1, Δ1-2
    <au>Gershenson, C</au>                             Δ1, Δ1-2
  </author>
  <journal>
    <issn>1094-7167</issn>                             Δ3a, Δ3-2
    <jtitle>IEEE INTELLIGENT SYSTEMS</jtitle>          Δ3a, Δ3-2
    <stitle>IEEE INTELL SYST</stitle>                  Δ3a, Δ3-2
  </journal>
  <enumeration>
    <year>2003</year>
    <volume>18</volume>
    <issue>4</issue>
    <spage>72</spage>
    <epage>75</epage>
  </enumeration>
</citation>
<reference>
  <source>
    <stitle>P 13 EUR M CYB SYST</stitle>               Δ3b, Δ2
    <author>BOLLEN, J</author>                         Δ2
    <spage>911</spage>                                 Δ2
    <year>1996</year>                                  Δ2
  </source>
</reference>
<reference>
  <source>
    <stitle>SWARM INTELLIGENCE</stitle>                Δ3b, Δ2
    <author>BONABEAU, E</author>                       Δ2
    <year>1998</year>                                  Δ2
  </source>
</reference>
    </didl:DIDL>
  </metadata>
</record>
```

Example 1: a resource's metadata record

## 2.2. The Three Layers of the LANL Multi-Relational Network and their Intra-Layer Projections

The proposed system is composed of three layers—an author-layer, a paper-layer, and a journal/proceedings-layer. Each layer is represented using a *co-authorship network*, a *citation network* or *co-citation network*, and a *journal reference network*, respectively. The goal of each network layer is to relate their respective nodes according to some homophilic property—some quality of similarity. For a multi-relational scholarly network, similarity comes by way of related research domains. The network construction algorithms described in this section associates the various resources of the repository either through *occurrence* or *co-occurrence* metadata.

Co-authorship networks define the past collaborations of authors in the community. When two authors publish a paper together, an edge between them in the co-authorship network can be created. The similarity between two co-authors rests on the intuitive notion that two researchers are similar in research ideas if they have collaborated on a paper together. The strength of this relationship can be determined in many ways. For one, the more two authors publish together, the greater their connection strength. If there are multiple co-authors on a particular paper, then the strength between any two co-authors is inversely proportional to the total amount of co-authors of the paper [Liu et al., 2005; Newman, 2001]. Thus, the weight between author $n_i$ and $n_j$ for all papers in the repository is defined as

$$w_{j,i}^{aut} = w_{i,j}^{aut} = \frac{\sum_{p \in P_{i,j}} \frac{1}{|A(p)|-1}}{|P_{i,j}|},$$

where $A(p)$ is the number of authors for paper $p$, $P_{i,j}$ is the set of all papers in the repository authored by both author $n_i$ and $n_j$, and $w_{i,j}^{aut}$ refers to the authorship metadata edge weight between authors $n_i$ and $n_j$. Minus one is used to set the upper limit of this association such that a paper with two authors has a weight a one and the sum is divided by the size of the set of all co-authored papers to ensure a value between 0 and 1. The connections within the co-authorship network have the semantic meaning: 'author X has co-authored with author Y'. Resource metadata similar to Example 1 contains the necessary information to build a co-authorship network—**Δ1**. In terms of Example 1, there are two authors, and assuming that no other papers have been co-authored by these authors, each direction's edge weight is 1.0.

The next layer above the co-authorship layer is the paper-layer and is composed of all paper, or manuscript, nodes. The paper-layer can be created using a citation network or a co-citation network. A citation network states that two papers are related if either one of the two papers references the other. Directed edges in a citation network represent a reference from one paper to another. The edge weight for a particular citation link is inversely proportional to the amount of references contained in the referencing paper. This is formally represented as

$$w_{i,j}^{cite} = \frac{1}{|C(i)|},$$

where $C(i)$ is the number of citations for paper $n_i$. In a citation network, a directed edge states that the first paper cites the second paper—**Δ2**. The two outgoing edges of the manuscript in Example 1 each have a weight of 0.5.

Co-citation is another method for constructing networks of related manuscripts. If two papers cite a third paper, then there exists some level of domain similarity between the two papers. The strength of tie between any two papers is calculated as,

$$w_{i,j}^{cocite} = w_{j,i}^{cocite} = \frac{|S(i,j)|}{|C(i)| + |C(j)| - |S(i,j)|}, \text{where } S(i,j) = \left\{ x \middle| x \in C(i) \wedge x \in C(j) \right\}$$

and $C(i)$ is the set of citations associated with paper $n_i$, $S(i,j)$ is the set of citations that are shared in common between papers $n_i$ and $n_j$, and the weight between any two papers is always between 0 and 1.

Finally, atop the paper layer is the journal and conference proceedings layer, but for brevity's sake, this layer will simply be called the journal-layer. The journal-layer is represented by a journal reference network. The journal reference network is constructed in a similar fashion to a paper citation network where a journal is similar to another journal if papers contained within it reference papers in the other. Formally,

$$w_{i,j}^{ref} = \frac{|R(i,j)|}{|R(i)|}$$

where $R(i,j)$ refers to the set of references that project from journal $n_i$ to journal $n_j$ and $R(i)$ refers to the set of all references of journal $n_i$. Again, the weight is between 0 and 1. The metadata record presented in Example 1 states that the paper, and its associated journal **Δ3a**, makes references to another journal **Δ3b**.

Table 1 outlines the edge semantics for each of the three networks described.

| Layer | Network Type | Intra-Edge Semantics |
|-------|--------------|----------------------|
| Δ1-author | co-authorship | 'author X has co-authored with author Y' |
| Δ2-paper | citation | 'paper X cites paper Y' |
| Δ3-journal | reference | 'journal X references journal Y' |

Table 1: the semantics of the intra-edge projections of each multi-relational network layer

## 2.3. The Three Layers of the LANL Multi-Relational Network and their Inter-Layer Projections

The previous subsection described the three layers of the scientific community's scholarly network and the meaning of their intra-layer connections. Each of these independent network layers can be interconnected with inter-layer projections to yield the full LANL multi-relational scholarly network. The author-layer projects to the paper-layer with edges that represent that an author has written a paper—**Δ1-2**. The paper-layer projects down to the author-layer to represent that a paper has been written by an author—**Δ1-2**. The paper-layer also projects up to the journal-layer representing that a paper has been published in a journal—**Δ3-2**. Finally, the journal-layer projects down to the paper-layer stating that a journal has published a paper—**Δ3-2**. Like each network layer, this information can also be derived given the resource's metadata record. The edge weights of these projects can be determined in an analogous fashion to the intra-layer projections.

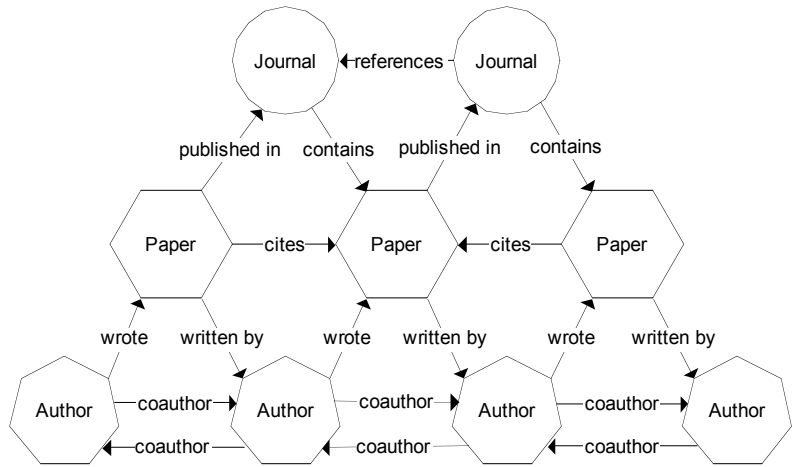

Figure 2: digital-library metadata can create the intra and inter-layer connections of LANL's multi-relational scholarly network

In Figure 2, upward projections are on the left hand side of the object and downward projections are on the right hand side. The imbalance of projections occurs in order to preserve the clarity of the diagram. Table 2 provides the semantics of the inter-network connections.

| Layer | Projections Down | Projections Up |
|---|---|---|
| Δ1-author | n/a | 'author X wrote paper Y' |
| Δ2-paper | 'paper X was written by author Y' | 'paper X is published in journal Y' |
| Δ3-journal | 'journal X contains paper Y' | n/a |

Table 2: the semantic meanings of the inter-layer connections within the multi-relational network

This section has described how a multi-relational network implementation of all the resources contained within the LANL repository can be constructed. Note that every repository represents its information in different formats, with potentially both metadata-rich and metadata-limited records. For this reason, no overarching multi-relational can be defined because it will differ for each repository. Furthermore, in the LANL network described here, there is nothing that prevents inter-layer connections between the author-layer to the journal-layer. This inter-layer projection would have the meaning that 'author X has published in journal Y' and, in reverse, 'journal X has published the paper of author Y'. Depending on the accuracy of the problem solving algorithm described next and the edge density of the network, such a connection may be appropriate. Other metadata currently exist. One particular type is usage data. Usage data defines the download patterns of resources in the repository and this information can be used to construct usage associations between paper and journal resources [Bollen et al., 2005]. Furthermore, link prediction algorithms can be employed to further increase the density of the network if need be [Liben-Nowell & Kleinberg, 2003].

# 3. Problem Solving in the Multi-Relational Network

In order to problem solve using the constructed multi-relational network, there must exist algorithms that utilizes the network to find solutions to particular problems. If multi-relational network is implemented within a hyper-text representation then a simple manual algorithm can occur. Researchers can move through the various resources in a web-surfing fashion until they find their appropriate resource. However, there are algorithms that can simulate this naturally occurring process. One such information retrieval algorithm is spreading activation [Cohen & Kjeldsen, 1987; Collins & Loftus, 1975; Crestani, 1997a; Crestani, 1997b; Crestani & Lee, 2000] or the random walker model of Markov Chain analysis. Spreading activation can simulate the way in which individuals browse a network, and can therefore, speed up the process by which resources in the network are discovered. Furthermore, other algorithms such as semantic queries or unique instance discovery algorithms can be implemented in the network to support the identification of unique resources and association paths [Boanerges & et. al. 2005][Lin & Chalupsky, 2003]. The point is that, depending on the individual's problem, different algorithms may be more appropriate than others. This paper will present one such algorithm that is a discrete relative-rank random walker algorithm.

In order to solve the five problems presented in the introduction to this paper, a non-converging random walker model is presented that is a discrete form of the spreading activation model. This algorithm is a relative-rank algorithm that ranks every node in an associative network relative to some initial set [Ziegler & Lausen, 2004]. The basic idea is to continuously map the problem representation (initial set of nodes) within and between the layers of the network until some desired format emerges (highly ranked nodes in a particular layer). The purpose of this section is to briefly describe how the discrete spreading activation algorithm takes a problem representation and transforms it into a solution by means of the multi-relational scholarly network substrate.

A random walker is an indivisible processing elements that represents the traversing behavior of scholar within the multi-relational scholarly network (e.g. perusing the citation network, the co-authorship network, etc.). The random walker begins its journey at particular node and traverses the edges of a network in a stochastic manner (i.e. randomly choosing edges biased by a larger weight) in order to identify the most highly visited nodes. In general, if the random walker starts at node $n_i$, then the node most visited after so many steps, is considered the most related node to $n_i$.

In order to simulate spreading activation and the idea the energy decays over time, a random walker has an energy value $\varepsilon_i \in [-1,1]$ and a decay value $\delta_i \in [0,1]$. A random walker's negative energy value represents an inhibitory action and thus, a depressing of a highly visited node. This will become important when identifying conflict of interest referees in the peer-review process. A random begins its journey at some source node and travels the network's edges depositing its current energy value at every node it traverses until its energy content has decayed below a given threshold, $\vartheta_i \in [0,1]$. The evolution of the random walker's energy value is defined by the decay function represented as

$$\varepsilon_i(t+1) = \begin{cases} (1-\partial)\varepsilon_i(t) & \text{when } |\varepsilon_i(t)| > \vartheta_i \\ 0 & \text{otherwise} \end{cases}$$

While a single random walker represents the search behavior of a single scholar, multiple random walkers can be used to explore more paths in the multi-relational scholarly network. The set P are the set of nodes that initially receive random walkers and $P_A$, $P_P$, and $P_J$ represent nodes in the author, paper, and journal-layer respectively, $P = P_A \cup P_P \cup P_J$. Furthermore a $P_X+$ is the set of random walkers in X-layer that are positive energy random walkers, $\varepsilon_i \in [0,1]$. Likewise, a $P_X-$ refers to a set of nodes in X-layer initially receiving negative energy random walkers, $\varepsilon_i \in [-1,0]$. Note that $P_X+$ and $P_X-$ are notation conventions and not a function on the set P. The initial random walker population propagates throughout the network, by way of the edge's connecting the various nodes, until all their energy has decayed below their energy threshold value. At the end of this process, any node in the network with positive energy is represented as the solution set, S, such that, $S_A$, $S_P$, and $S_J$ represent the various solution's in author, paper, and journal format, $S = S_A \cup S_P \cup S_J$. Depending on the stage of the scholarly communication process, the initial distribution of random walkers have a different meaning. In some cases, it is important to identify papers, sometimes journals, and sometimes scholars. The next section will walk through an example of the proposed systems search capabilities.

# 4. The LANL Multi-Relational Network and Five Problem Solving Processes

The specific LANL multi-relational network described in Section 2 and the relative-rank random walker algorithm described in Section 3 can support a researcher through the development of an initial idea to the distribution of the idea's manuscript to interested members in the community. Each of the five major problem solving endeavors discussed in Section 1 can be addressed using the random walker problem/solution pattern-matching algorithm within the multi-relational scholarly network. The following subsections discuss the five stages of the scholarly communication process enumerated in the introduction of this paper.

## 4.1. Finding References for an Inspiration

Formal papers, published in journals or proceedings, are the artifacts of scholarly communication and their production relies on the efforts of an author and the previously published research of others in the community. Every paper begins with an initial inspiration that grows into a formalized idea ready for distribution amongst the community. In many cases, that idea is triggered by some inspirational paper that the researcher has read. This initial paper is called a keystone paper. From here, a novel idea must be nurtured by and tied to other papers in the domain. So, from an initial idea, sparked by some keystone paper, the researcher must find the niche for this new, soon to be formalized, idea. Therefore, the problem is to find a set of papers that are related to the researchers discovered keystone paper.

For this researcher's problem, the $P_P+$ set will contain only the keystone paper. These positive energy random walkers propagate within the paper-layer and between the author and journal-layers incrementing the energy of all the nodes they pass through. For example, the random walkers at $t=1$ will be solely contained at the node representing the keystone paper. At $t=2$, the journal that published the keystone paper, the references of the keystone paper, and the authors of the keystone paper will have random walkers. Over time, the initial energy distribution diffuses over the network exposing all resources related to the initial keystone paper. At the end of the random walker algorithm there is an $S_A$, $S_P$, and an $S_J$ set. Since the researcher is concerned with finding potential references for the new idea, only the top ranked nodes of the $S_P$ set are reviewed.

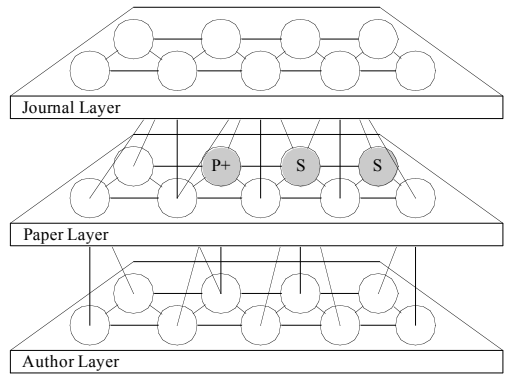

Figure 1: $P_P+$ (keystone paper) $\rightarrow$ $S_P$ (related papers)

The researcher has utilized the system to match the problem—find papers related to the keystone paper—to a solution—a collection of related papers. Furthermore, the researcher can trim the returned $S_p$ set and use that filtered set as a $P_P+$ set to again stimulate the multi-relational network to search for yet more related publications.

## 4.2. Finding Collaborators to Progress the Idea

Now that the researcher has a collection of papers that are representative of the initial idea, the researcher may be interested in finding potential collaborators. A way to make this possible via the proposed system is to locate a group of researchers that are related in thought to the collection of papers retrieved in Section 4.1. The intuitive notion is that individuals whom are related in thought to the papers derived in Section 4.1 are more than likely going to understand the author's vague notion of an idea and, in turn, be able to provide assistance in formalizing the idea more clearly. Furthermore, if the researcher's idea refers to multiple domains, then the stimulation of all the associated papers will provide more energy to those individuals whose research interests cover the full range of domains. To determine collaborators, the random walker input node set, $P_P+$, is the solution set, $S_P$, from Section 4.1.

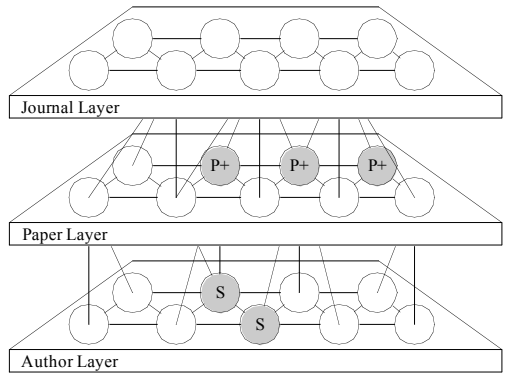

Figure 2: P$_P$+ (related papers) → S$_A$ (collaborators)

These positive energy random walkers propagate throughout the multi-relational network—up to the journal-layer and down to the author-layer. Since the researcher is interested in finding collaborating authors, the researcher reviews the S$_A$ solution-set. The highly ranked authors in S$_A$ can then be contacted for discussion and a potential collaborative paper writing relationship may form.

## 4.3. Finding a Journal for the Formalized Idea

After finding papers to reference and scholars to collaborate with, the original researcher and his collaborators have written a paper that contains a novel idea that they believe is worth publishing in a journal. In order to find a good arena to publish their work, the authors must find a journal that is related to the written paper. There could be many ways to represent this problem, but for the example demonstrated in Figure 3, the problem is represented as a set of nodes in both the paper and author layers. P$_A$+ is the set of all collaborating authors on the new paper. In the paper-layer, P$_P$+ is the set of all papers referenced by the new unpublished paper—the papers in the bibliography of the newly written pre-print.

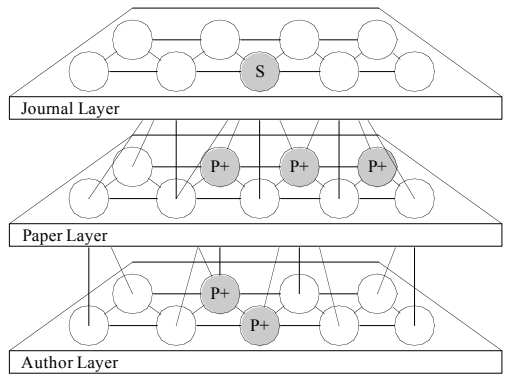

Figure 3: P$_P$+ (referenced papers) · P$_A$+ (authors) → S$_J$ (submitting journal)

Since the researchers are interested in a potential journal to publish in, the highest energy nodes of the $S_J$ set are reviewed as potential publishing venues. The intuitive idea is to search the network for journals that have published the papers of the cited articles and have published papers of the collaborating scientists and their co-authors. It is possible to organize the initial distribution of random walkers differently. For example, one other method could be to stimulate the journals and authors of the referenced papers. Such a stimulation pattern will return a list of journals related to the journals of the paper's bibliography convoluted with the list of journals that are related to the authors of the bibliography. Depending on the results of $S_J$, a different problem representation may be more appropriate.

## 4.4.  Finding Peer-Reviewers to Certify the Pre-print

At this stage, the context of the scholarly communication process has shifted from the perspective of a collaborating team of scholars to the editors of a journal. With the paper written and submitted, it is up to the editors to certify the quality of the manuscript. This is usually accomplished by means of the peer-review process. The editors of the journal locate individuals within the community for whom they believe to be experts in the paper's domain. The paper is then distributed to these referees for review. Finding experts in the field can be a laborious task requiring manual effort on behalf of the journal editor. The proposed system can solve this problem by matching an unpublished paper to a set of expert individuals in the field. The unpublished paper is abstractly represented in a $P_P+$ set as all its bibliographic references since its representative node has yet to be placed in the network—the paper has yet to be published. Secondly, a $P_J+$ set can be created to contain only the reviewing journal. The reason for stimulating the reviewing journal is that the reviewing journal may want to search the web of authors whom have published in their journal previous. These authors understand the quality requirements of the journal and may be more inclined to review the paper. Finally, and of most interest is the $P_A-$ set. $P_A-$ is the set of authors whom have written the paper. Since an author cannot be a referee to their own paper, and since it is advisable that past collaborators of those authors also not be referees due to conflict of interest situations, these nodes should be given negative energy to inhibit the activation of themselves and the individuals around them in the co-authorship network (Rodriguez et al., 2006).

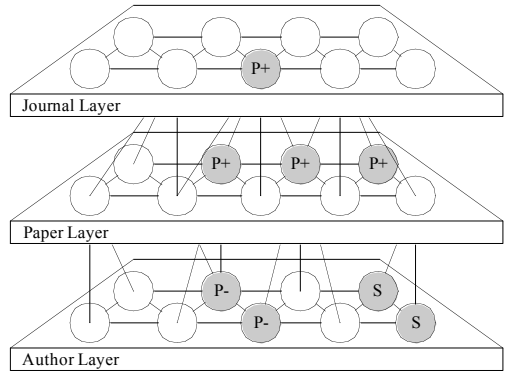

Figure 4: P$_J$+ (reviewing journal) · P$_P$+ (referenced papers) · P$_A$- (authors) → S$_A$ (reviewers)

The process of using a co-authorship network and a radom walker algorithm to find peer-reviewers for a pre-print has been applied in (Rodriguez et al., 2006; Rodriguez et al., 2007) with positive results. Furthermore, the set S$_A$ need not be trimmed to only the top 3 or 4 individual referees, but in fact, each scholar in S$_A$ can be deemed a peer-reviewer and their decisions on the review process can be weighted. Each member in that set, after the algorithm is complete, has some proportion of the total network energy. This varying degree of energy can be used to determine the individual's decision-making influence in the review process—where more expert individuals, with regards to the paper in question, should have more influence in the accept/reject decision of the paper (i.e. open commentary peer-review). This idea has been proposed as a means of augmenting digital-library technology to support the peer-review process and is further developed in (Rodriguez et al., 2006).

## 4.5.   Finding Related Scholars to Read the Manuscript

Finally, the paper has been written, submitted, and accepted. It is the role of the journal to find the largest distribution-base for the paper. Journals can aid the author and the community by matching the paper to a set of readers who may be interested in the newly published paper. This is done by creating a P$_P$+ set containing the newly published paper's bibliographic references and creating a P$_A$+ set containing the authors of the paper. The random walker algorithm runs and the solution set, S$_A$, is the set of all individuals in the author layer whom are related in thought to the newly published manuscript. If these individuals have provided email address with the digital-library system, then these authors can be automatically contacted with an email containing the title of the paper, its abstract, and a URL to the paper in the digital-library repository (c.f. ArXiv). Furthermore, a P$_J$+ could be used to provide energy to authors whom have published in the journal previous. Again, depending on the specific implementation, different problem representations will be more appropriate than others.

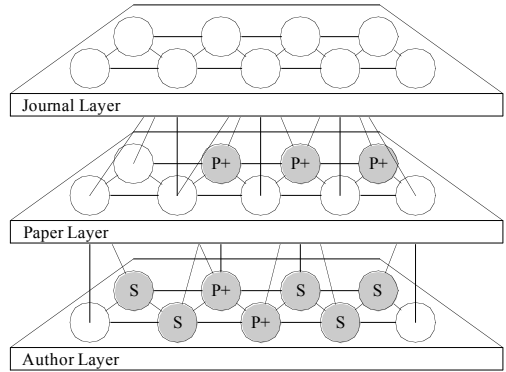

Figure 5: $P_A$+ (paper authors) $\cdot$ $P_P$+ (referenced papers) $\rightarrow$ $S_A$ (interested readers)

| Scholarly Process | Problem Representation | Solution |
|---|---|---|
| References | $P_P$+ (keystone paper) | $S_P$ (related papers) |
| Collaborators | $P_P$+ (related papers) | $S_A$ (potential collaborators) |
| Journals | $P_P$+ (related papers) $\cdot$ $P_A$+ (authors) | $S_J$ (submitting journal) |
| Peer-reviewers | $P_J$+ (journal) $\cdot$ $P_P$+ (related papers) $\cdot$ $P_A$- (authors) | $S_A$ (potential reviewers) |
| Readers | $P_P$+ (related papers) $\cdot$ $P_A$+ (authors) | $S_A$ (interested readers) |

Table 3: the problem and solution representations of 5 problem solving processes
associated with scholarly communication

A single multi-relational netowrk can provide the functionality to perform many complex problem solving tasks for the community (Table 3). This section only addressed five common problems encountered in the scholarly communication process. If there exists a problem that can be represented as a set of authors, papers, and journals, and there exists a complementary solution within the same space, then the proposed system can be used as a medium for such problem solving. It is important to realize the different problem representations could potentially return different solutions (Heylighen, 1988). Furthermore, more advanced random walker algorithms can be employed such as the grammar-based random walker model for semantic networks (Rodriguez, 2007). The demonstration of these techniques is out of the scope of this paper and is left to a future publication with a more thorough analysis of the proposed system.

# 5. Conclusion

Thanks in part to digital-library repositories and the OAI-PMH protocol; what has emerged in the scientific community is a computational medium that supports collective problem solving. Each individual stigmergetically (Holland & Melhuish, 1999) contributes to the evolution of this structure through their publication efforts and, in doing so, implicitly provides the intelligent pathways necessary for automated problem solving. Many digital-library repositories maintain rich metadata records that can be exploited in novel ways. Currently, many repositories only support keyword searching or citation network surfing. The purpose of this paper was to present a novel use of repository metadata as a complementary search service for use by scholars in the community.

The LANL multi-relational network proposed in this paper is one type of network that can be created from the metadata associated with LANL's repository resources. There are still many other problem's that can be developed that have not been formalized by this paper. For instance, other network layers can be instantiated such as research institutions and funding sources. Such layers could provide the necessary infrastructure to determine the most appropriate individual for a research position or the most appropriate institution to receive a research grant. In this way, a multi-relational network can encompass many more issues associated with the scholarly process. And still, outside the domain of science, where rich metadata information may exist, these ideas could easily be mapped over. To conclude, future research into multi-relational network and associated network analysis algorithms can be used to provide digital-library's more tools to support the scholarly communication process.

# 6. Acknowledgements

These ideas were initially developed at Free University of Brussels in Belgium during discussions with Francis Heylighen. Johan Bollena and Thad Paulson edited drafts of this paper. This research was supported by a GAANN Fellowship from the U.S. Department of Education, the Fonds voor Wetenschappelijk Onderzoek – Vlaanderen, and the Los Alamos National Laboratory.